\documentclass[twocolumn,showpacs,preprintnumbers,amsmath,amssymb]{revtex4}
\input psfig.sty

\usepackage{graphicx}
\usepackage{dcolumn}
\usepackage{bm}

\begin{document}

\title{Cosmological constraints on Chaplygin gas dark energy from galaxy clusters X-ray 
and supernova data}

\author{J. V. Cunha$^{1}$} \email{jvital@dfte.ufrn.br}

\author{J. S. Alcaniz$^{1}$} \email{alcaniz@dfte.ufrn.br}

\author{J. A. S. Lima$^{1,2}$} \email{limajas@dfte.ufrn.br}

\affiliation{$^{1}$Departamento de F\'{\i}sica, Universidade Federal do
Rio Grande do Norte C.P. 1641, Natal - RN, 59072-970, Brazil\\$^{2}$Instituto de
Astronomia, Geof\'{\i}sica e Ci\^encias Atmosf\'ericas,
Universidade de S\~ao Paulo, Rua do Mat\~ao, 1226 - Cidade Universit\'aria, 05508-900, 
S\~ao Paulo - SP, Brasil}

\date{\today}

\begin{abstract}

The recent observational evidences for the present accelerated
stage of the Universe have stimulated renewed interest for
alternative cosmologies. In general, such models contain an
unknown negative-pressure dark component that explains the
supernova results and reconciles the inflationary flatness
prediction ($\Omega_{\rm{T}} = 1$) and the cosmic microwave
background measurements with the dynamical estimates of the
quantity of matter in the Universe ($\Omega_{\rm{m}} \simeq 0.3
\pm 0.1$). In this paper we study some observational consequences
of a dark energy candidate, the so-called generalized Chaplygin
gas which is characterized by an equation of state $p_{C} =
-A/\rho_{C}^{\alpha}$, where $A$ and $\alpha$ are positive
constants. We investigate the prospects for constraining the
equation of state of this dark energy component by combining
Chandra observations of the X-ray luminosity of galaxy clusters,
independent measurements of the baryonic matter density, the
latest measurements of the Hubble parameter as given by the HST
Key Project and data of the Supernova Cosmology Project. We show
that very stringent constraints on the model parameters can be
obtained from this combination of observational data.

\end{abstract}

\maketitle

\section{Introduction}

One of the most important goals of current cosmological studies is
to unveil the nature of the so-called dark energy or {\emph
{quintessence}}, the exotic negative-pressure component
responsible for the accelerating expansion of our Universe. Over
the last years, a number of candidates for this dark energy have
been proposed in the literature \cite{darkenergy}, with the vacuum
energy density (or cosmological constant) and a dynamical scalar
field \cite{peebles,cald} apparently constituting the most plausible
explanations. From the observational viewpoint, these two class of
models are currently considered our best description of the
observed Universe whereas from the theoretical viewpoint they
usually face fine-tunning problems, notably the so-called
cosmological constant problem \cite{wein} as well as the cosmic
coincidence problem, i.e., the question of explaining why the
vacuum energy or the scalar field dominates the Universe only very
recently. The later problem happens even for tracker versions of
scalar field models in which the evolution of the dark energy
density is fairly independent of initial conditions \cite{peebles,cald1}.

Among the many dark energy candidates, a recent and very
interesting proposal has been suggested by Kamenshchik {\it et
al.} \cite{kamen} and developed by Bili\'c {\it et al.}
\cite{bilic} and Bento {\it et al.} \cite{bento}. It refers to the
so-called Chaplygin gas (C), an exotic fluid whose equation of
state is given by
\begin{equation}
p_{C} = -A/\rho_{C}^{\alpha},
\end{equation}
with $\alpha = 1$ and $A$ a positive constant. Actually, the above
equation for $\alpha \neq 1$ generalizes the original Chaplygin
equation of state proposed in Ref. \cite{bento} whereas for
$\alpha = 0$, the model behaves like scenarios with cold dark
matter plus a cosmological constant ($\Lambda$CDM).

In the context of the Friedman-Robertson-Walker (FRW) cosmologies,
whether one inserts Eq. (1) into the energy conservation law
($u_{\mu}T^{{\mu}{\nu}}_{;\nu} = 0$), the following expression for
the energy density is immediately obtained
\begin{equation} \label{limB}
\rho_{C} = \left[A + B\left(\frac{R_o}{R}\right)^{3(1 +
\alpha)}\right]^{\frac{1}{1 + \alpha}},
\end{equation}
or, equivalently,
\begin{equation}
\rho_{C} = \rho_{C_{o}}\left[A_s + (1 -
A_s)\left(\frac{R_o}{R}\right)^{3(1 + \alpha)}\right]^{\frac{1}{1
+ \alpha}},
\end{equation}
where $\rho_{C_{o}}$ is the current energy density (from now on a
subscript $o$ means present day quantities, and C denotes either
the Chaplygin gas or its generalized version). The function $R(t)$
is the cosmic scale factor, $B = \rho_{C_{o}}^{1 + \alpha} - A$ is
a constant and $A_s = A/\rho_{C_{o}}^{1 + \alpha}$ is a quantity
related to the present day Chaplygin adiabatic sound speed
($v_s^{2} = \alpha A/\rho_{C_{o}^{1 + \alpha}}$). As can be seen
from the above equations, the C-gas interpolates between
non-relativistic matter ($\rho_{C}(R \rightarrow 0) \simeq
\sqrt{B}/R^{3}$) and negative-pressure dark component regimes
($\rho_{C}(R \rightarrow \infty) \simeq \sqrt{A}$). This
particular behavior of the Chaplygin gas inspired some authors to
propose a unified scheme for the cosmological ``dark sector"
\cite{bilic,bento,jailson}, an interesting idea which has also
been considered in many different contexts \cite{unif} (see,
however, \cite{teg}).

In the theoretical front, a connection between the Chaplygin
equation of state and string theory had long been identified by
Bordemann and Hoppe \cite{bord} and Hoppe \cite{hop} (see also
\cite{jack} for a detailed review). As explained in such
references, a Chaplygin gas-type equation of state is associated
with a parametric description of the invariant Nambu-Goto
$d$-brane action in a $d + 2$ spacetime. In the light-cone
parameterization, such an action reduces itself to the action of a
Newtonian fluid which obeys Eq. (1) with $\alpha = 1$, with the
C-gas corresponding effectively to a $d$-branes gas in a ($d +
2$)-dimensional spacetime.

Another interesting connection is related to recent attempts of
describing the dark energy component through the original
Chaplygin gas or its generalized version. Such a possibility has
provoked a growing interest for exploring the observational
consequences of this fluid in the cosmological context. For
example, Fabris {\it et al.} \cite{fabris} analyzed some
consequences of such scenario using type Ia supernovae data (SNe
Ia). Their results indicate that a Universe completely dominated
by the Chaplygin gas is favored when compared with $\Lambda$CDM
models. Recently, Avelino {\it et al.} \cite{avelino} used
a larger sample of SNe Ia and the shape of the matter power
spectrum to show that such data restrict the model to a behaviour
that closely matches that of a $\Lambda$CDM models while Bento
{\it et al.} \cite{bento1,bert1} showed that the location of the
CMB peaks imposes tight constraints on the free parameters of the
model. More recently, Dev, Alcaniz \& Jain \cite{dev} and Alcaniz,
Jain \& Dev \cite{jailson} investigated the constraints on the
C-gas equation of state from strong lensing statistics and
high-$z$ age estimates, respectively, while Silva \& Bertolami
\cite{bert} studied the use of future SNAP data together with the
result of searches for strong gravitational lenses in future large
quasar surveys to constrain C-gas models. Makler {\it et al.} \cite{MI} also showed that 
such models are consistent with current SNe Ia data for a broad range of parameters. 
The trajectories of
statefinder parameters \cite{stat} in this class of scenarios were
studied in Ref. \cite{stat1} while constraints involving Cosmic
Microwave Background (CMB) data have also been extensively
discussed by many authors \cite{bento1,bert1,fin,finelli}.

In this work we study the possibility of constraining the
generalized Chaplygin equation of state from X-ray luminosity of
galaxy clusters. With basis on measurements of the mean baryonic
mass fraction in clusters as a function of redshift, we consider
the method originally proposed by Sasaki \cite{sasa} and Pen
\cite{pen}, and further modified by Allen {\it et al.}
\cite{allen,allen1} who analyzed the X-ray observations in some
relaxed lensing clusters observed with Chandra in the redshift
interval $0.1 < z < 0.5$ (see also \cite{ett}). By inferring the
corresponding gas mass fraction, Allen and collaborators placed
observational limits on the total matter density parameter,
$\Omega_{\rm m}$, as well as on the density parameter
$\Omega_{\Lambda}$, associated to the vacuum energy density. More
recently, a similar analysis has also been applied to conventional
quintessence models with an equation of state $p_x = \omega
\rho_x$ by Lima {\it et al.} \cite{lima}.

The paper is organized as follows. In Sec. II we present the field
equations and distance formulas necessary to our analysis. In Sec.
III the corresponding limits on C-gas models from X-ray luminosity
of galaxy clusters are derived. We also examine the limits from a
statistical combination between X-ray data and recent SNe Ia
observations. Finally, in Sec. IV, we finish the paper by
summarizing the main results and comparing our constraints with
others derived from independent analyses.

\section{The Chaplygin gas model}

The FRW equation for a spatially flat, homogeneous, and isotropic
scenarios driven by nonrelativistic matter and a separately
conserved C-gas component reads
\begin{eqnarray}
(\frac{{\dot R}}{R})^{2} & = & H_o^{2}
\{\Omega_{m}(\frac{R_o}{R})^{3} + \\ \nonumber & & + (1 -
\Omega_{m})[A_s + (1 - A_s) (\frac{R_o}{R})^{3(\alpha +
1)}]^{\frac{1}{\alpha + 1}}\},
\end{eqnarray}
where an overdot denotes time derivative, $H_{o} = 100h
{\rm{Km.s^{-1}.Mpc^{-1}}}$ is the present value of the Hubble
parameter, $\Omega_m$ is the matter density parameter, and the
dependence of the C-gas energy density with the scale factor comes
from Eq. (3).

The comoving distance $r_1(z)$ to a light source located at $r =
r_1$ and $t = t_1$ and observed at $r = 0$ and $t = t_o$ is given
by
\begin{equation}
r_1(z) = \frac{1}{R_oH_o}\int_{x'}^{1}\frac{dx}{x^{2}{\cal{F}}(x,
\Omega_m, A_s, \alpha)}
\end{equation}
where $x' = R(t)/R_o = (1 + z)^{-1}$ is a convenient integration
variable and the dimensionless function ${\cal{F}}(x, \Omega_m,
A_s, \alpha)$ is given by
\begin{eqnarray}
{\cal{F}} & = & \left[\Omega_{m}x^{-3} + (1 - \Omega_{m})\left(A_s
+ \frac{(1 - A_s)}{x^{3(\alpha + 1)}}\right)^ {\frac{1}{\alpha +
1}}\right]^{1/2}.
\end{eqnarray}
Now, in order to derive the constraints from X-ray gas mass
fraction on the C-gas, let us consider the concept of angular
diameter distance, $D_A(z)$. Such a quantity is defined as the
ratio of the source diameter to its angular diameter, i.e.,
\begin{equation}
D_{A} = {\ell \over \theta} = R(t_1)r_1 = (1 + z)^{-1}R_{o}r_1(z),
\end{equation}
which provides, when combined with Eq. (5),
\begin{equation}
D_{\rm A}^{\rm{C}} = \frac{H_o^{-1}}{(1 + z)}\int_{x'}^{1} {dx
\over x^{2} {\cal{F}}(x, \Omega_m, A_s, \alpha)}.
\end{equation}
As one may check, for $A_s = 0$ and $\alpha = 1$ the above
expressions reduce to the standard cold dark matter model (SCDM).
In this case, the angular diameter distance can be written as
\begin{eqnarray}
D_{\rm A}^{\rm{SCDM}} = \frac{2H_o^{-1}}{(1 + z)^{3/2}}\left[(1 +
z)^{1/2} - 1\right].
\end{eqnarray}

\section{Limits from x-ray gas mass fraction}

Following Allen {\it et al.} \cite{allen, allen1} and Lima {\it et
al.} \cite{lima}, we consider the Chandra data consisting of six
clusters distributed over the redshift interval $0.1 < z < 0.5$.
The data are constituted of regular, relatively relaxed systems
for which independent confirmation of the matter density parameter
results is available from gravitational lensing studies. The X-ray
gas mass fraction ($f_{\rm gas}$) values were determined for a
canonical radius $r_{2500}$, which is defined as the radius within
which the mean mass density is 2500 times the critical density of
the Universe at the redshift of the cluster. In order to generate
the data set the SCDM model with $H_o = 50
{\rm{Km.s^{-1}.Mpc^{-1}}}$ was used as the default cosmology (see
\cite{allen} for details).

\begin{figure}
\centerline{\psfig{figure=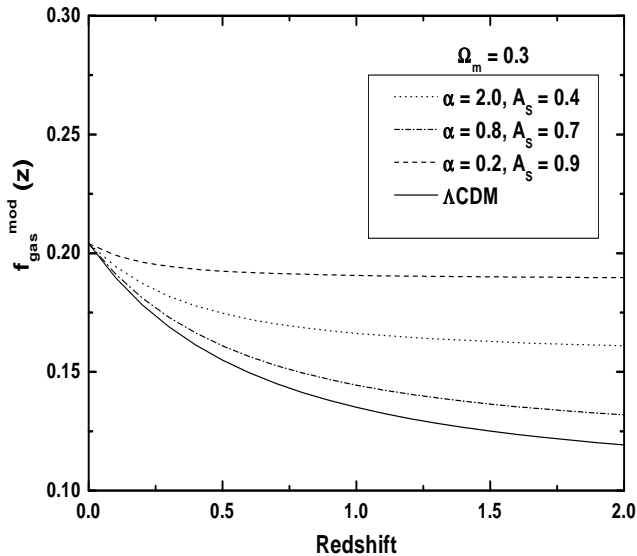,width=3.4truein,height=3.0truein}
\hskip 0.1in} \caption{The model function $f_{\rm gas}^{\rm mod}$
as a function of the redshift for selected values of $A_s$ and
$\alpha$ and fixed values of $\Omega_{\rm{m}} = 0.3$, $\Omega_{\rm
b}h^{2} = 0.0205$ and $h = 0.72$.}
\end{figure}

By assuming that the baryonic mass fraction in galaxy clusters
provides a fair sample of the distribution of baryons at large
scale (see, for instance, \cite{hogan}) and that $f_{\rm gas}
\propto D_{\rm{A}}^{3/2}$ \cite{sasa}, the model function is
defined as \cite{allen}
\begin{eqnarray}
f_{\rm gas}^{\rm mod}(z_{\rm i}) = \frac{ b\Omega_{\rm b}}
{\left(1+0.19{h}^{3/2}\right) \Omega_{\rm m}} \left[ 2h \,
\frac{D_{\rm A}^{\rm{SCDM}}(z_{\rm i})}{D_{\rm A}^{\rm{C}}(z_{\rm
i})} \right]^{1.5},
\end{eqnarray}
where the bias factor $b \simeq 0.93$ \cite{bia} is a parameter
motivated by gas dynamical simulations that takes into account the
fact that the baryon fraction in clusters is slightly depressed
with respect to the Universe as a whole \cite{frenk}. The term
$(2h)^{3/2}$ represents the change in the Hubble parameter between
the default cosmology and quintessence scenarios and the ratio
${D_{\rm A}^{\rm{SCDM}}(z_{\rm i})}/{D_{\rm A}^{\rm{C}}(z_{\rm
i})}$ accounts for deviations in the geometry of the universe from
the default cosmology (SCDM model). In Fig. 1 we show the behavior
of $f_{\rm gas}^{\rm mod}$ as a function of the redshift for some
selected values of $A_s$ and $\alpha$ having the values of
$\Omega_{\rm b}$ and $h$ fixed. For the sake of comparison, the
current favored cosmological model, namely, a flat scenario with
$70\%$ of the critical energy density dominated by a cosmological
constant ($\Lambda$CDM) is also shown. In order to have
bidimensional plots we fix the value of $\Omega_{\rm m}$ as
suggested by dynamical estimates \cite{calb}, i.e., 0.3 in Fig. 1
as well as in all statistical analyses involving the generalized
Chaplygin gas. However, in the case of a conventional C-gas
($\alpha = 1$), the density parameter $\Omega_{\rm m}$ has also
been considered a free parameter to be adjusted by the data.

The cosmological parameters $A_s$ and $\alpha$ are determined by
using a $\chi^{2}$ minimization with the priors $\Omega_{\rm
b}h^{2} = 0.0205 \pm 0.0018$ \cite{nuc} and $h = 0.72 \pm 0.08$
\cite{freedman} for the range of $A_s$ and $\alpha$ spanning the
interval [0,1] in steps of 0.02,
\begin{eqnarray}
\chi^2 & = &\sum_{i = 1}^{6} \frac{\left[f_{\rm gas}^{\rm
mod}(z_{\rm i}, \Omega_{\rm{m}}, A_s, \alpha)- f_{\rm gas,\,i}
\right]^2}{\sigma_{f_{\rm gas,\,i}}^2} + \\ \nonumber & & \quad
\quad + \left[\frac{\Omega_{\rm{b}}h^{2} -
0.0205}{0.0018}\right]^{2} + \left[\frac{h -
0.72}{0.08}\right]^{2},
\end{eqnarray}
where $\sigma_{f_{\rm gas,\,i}}$ are the symmetric
root-mean-square errors for the SCDM data. The $68.3\%$ and
$95.4\%$ confidence levels are defined by the conventional
two-parameters $\chi^{2}$ levels 2.30 and 6.17, respectively.

\begin{figure}
\centerline{\psfig{figure=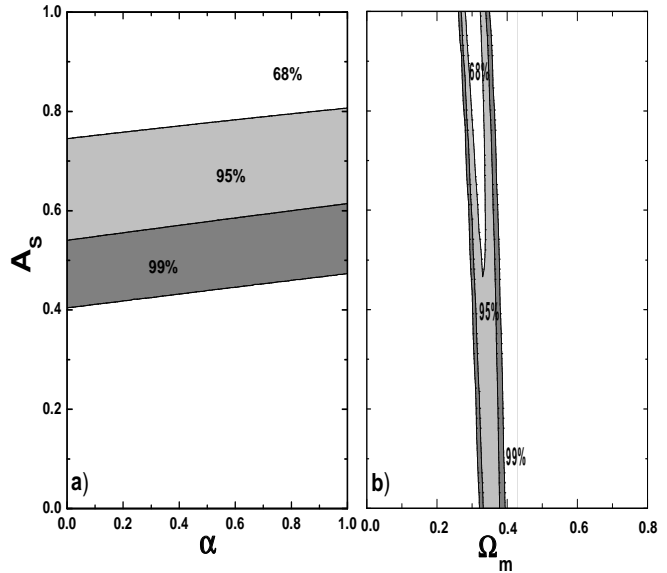,width=3.4truein,height=3.0truein}
\hskip 0.1in} \caption{{\bf{a)}} The $\Delta \chi^{2}$ contours
for the $\alpha - A_s$ plane according to the X-ray data discussed
in the text. The contours correspond to 68\%, 95\% and 99\%
confidence levels. The value of the matter density parameter has
been fixed at $\Omega_{\rm{m}} = 0.3$. {\bf{b)}} $\Omega_{\rm{m}}
- A_s$ plane for the original C-gas model ($\alpha = 1$). At
95.4\% we find $\Omega_{\rm{m}} = 0.3 \pm 0.02$ while the entire
range of $A_s$ is allowed. Note that the X-ray data constrain tightly the matter density
parameter.}
\end{figure}

In Fig. 2a we show contours of constant likelihood (68\%,
95\% and 99\%) in the parameter space $\alpha - A_s$ for the X-ray data
discussed earlier. From the above equation we find that the best
fit model occurs for $A_s = 1$ which, according to Eq. (4), is independent of the index
$\alpha$ and equivalent to a $\Lambda$CDM universe. Such model corresponds to a
accelerating scenario with the deceleration
parameter $q_o = -0.55$ \footnote{Note that for $A_s = 1$, Eq. (4) does not depend on
the parameter $\alpha$. Therefore, the smoothness of the curves at these points is a
consequence of the step used for the parameters in the code.}. From this figure, we
also see that both $A_s$ and $\alpha$ are quite insensitive to these data and that, at
95.4\%
c.l., one can limit the parameter $A_s$ to be $> 0.52$. Figure
2b shows the plane $\Omega_{\rm{m}} - A_s$ for the conventional
C-gas ($\alpha = 1$). As one should expect from different analyses
\cite{allen,lima}, the matter density parameter is very well
constrained by this data set while the parameter $A_s$ keeps quite
insensitive to it. The best fit occurs for models lying in the
interval $A_s = [0,1]$ and $\Omega_{\rm{m}} = 0.3$. At 95.4\%
c.l., we find $0.268 \leq \Omega_{\rm{m}} \leq 0.379$. For a X-ray analysis where the Cg
plays the role of a unified model for dark matter/energy, see \cite{ioav1}.

\begin{figure}
\centerline{\psfig{figure=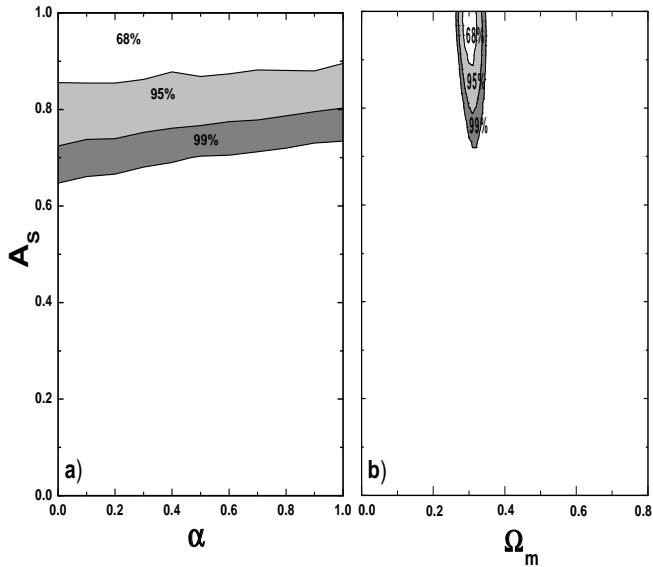,width=3.4truein,height=3.0truein}
\hskip 0.1in} \caption{{\bf{a)}} The likelihood contours in the
$\alpha - A_s$ plane for the joint X-ray + SNe Ia analysis
described in the text. The contours correspond to 68\%, 95\% and
99\% confidence levels. {\bf{b)}} The $\Omega_{\rm{m}} - A_s$
plane for the joint X-ray + SNe Ia analysis. The best fit values
are located at $A_s = 0.98$ and $\Omega_{\rm{m}} = 0.3$ At 95.4\%
we find $A_s \geq 0.84$ and $0.273 \leq \Omega_{\rm{m}} \leq
0.329$.}
\end{figure}

\subsection{Joint analysis with SNe Ia}

By combining the X-ray and SNe Ia data sets, more stringent
constraints on the cosmological parameters $\Omega_{\rm{m}}$ and
$A_s$ are obtained. As it was shown elsewhere, the parameter
$\alpha$ is highly insensitive to the SNe Ia data. To perform such
analysis, we follow the conventional magnitude-redshift test (see,
for example, \cite{sneA}) and use the SNe Ia data set that
corresponds to the primary fit C of Perlmutter {\it et al.}
\cite{perlmutter} together with the highest supernova observed so
far, i.e, the 1997ff at $z = 1.755$ and effective magnitude
$m^{eff} = 26.02 \pm 0.34$ \cite{riess} and two newly discovered
SNe Ia, namely, SN 2002dc at $z = 0.475$ and $m^{eff} =22.73 \pm
0.23$ and SN 2002dd at $z = 0.95$ and $m^{eff} = 24.68 \pm 0.2$
\cite{blak}. Figures 3a, 3b and 4 show the results of our
analysis. In Fig. 3a we display contours of the combined
likelihood analysis for the parametric space $A_s - \alpha$. In
comparison with Fig. 2a we see that the available parameter space
is reasonably modified with the value of $A_s$ constrained to be
$> 0.73$ at 95.4\% c.l. and the entire interval of $\alpha =
[0,1]$ allowed. The best fit model occurs for values of $A_s =
0.98$ and $\alpha = 0.93$ with $\chi^{2}_{min} = 61.38$ and $\nu =
61$ degrees of freedom ($\chi^{2}_{min}/\nu \simeq 1.0$). The most
restrictive limits from this joint analysis are obtained for the
original version of C-gas ($\alpha = 1$). In this case, the plane
$\Omega_{\rm{m}} - A_s$ (Fig. 3b) is tightly constrained with the
best fit values located at $A_s = 0.98$ and $\Omega_{\rm{m}} =
0.3$ with $\chi^{2}_{min}/\nu \simeq 1.0$. At 95.4\% this analysis
also provides $A_s \geq 0.84$ and $0.273 \leq \Omega_{\rm{m}} \leq
0.329$.  Note that the contours $\alpha-A_s$ (Fig. 3a) and $\Omega_m - A_s$ (Fig. 2b)
are almost orthogonal, thereby explaning the shape 
of the $\Omega_m - A_s$ plane appearing in Fig. 3b. We also observe that by extending
the $\alpha-A_s$ plane
to the interval [0,2] the new best fit values ($A_s = 1.02$ and
$\alpha = 0.45$), although completely modified in comparison with
the previous ones, are still in agreement with the causality ($A_s
\leq 1/\alpha$) imposed by the fact that the adiabatic sound speed
$v_s^{2} = dp/d\rho$ in the medium must be lesser than or equal to
the light velocity (see Eq. 1). Some basic results of the above
analysis are displayed in Fig. 4.

\begin{figure}
\centerline{\psfig{figure=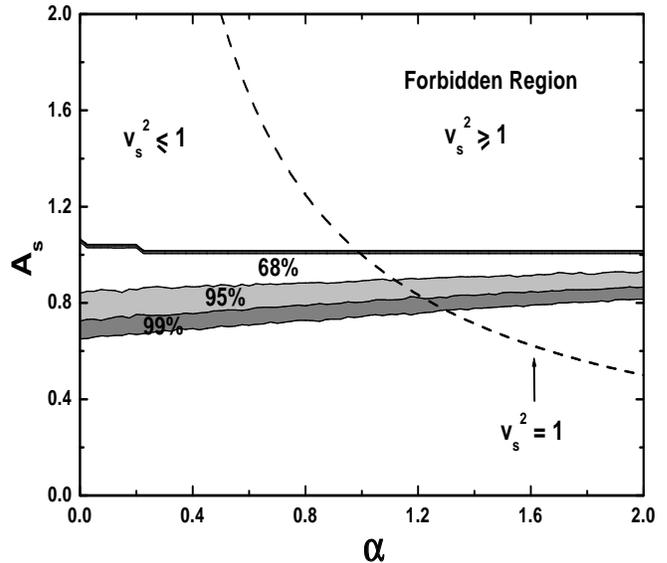,width=3.4truein,height=3.0truein}
\hskip 0.1in} \caption{The extended $A_s - \alpha$ plane for the
joint X-ray + SNe Ia analysis. Although completely modified in
comparison with the previous ones, the best fit values for this
extended analysis ($A_s = 1.02$ and $\alpha = 0.45$) are still in agreement with the
causality
imposed by the sound/light velocity ($A_s \leq 1/\alpha$). The
dashed hyperbola corresponds to the limit condition $v_s^{2}=1$.}
\end{figure}

\section{Discussion and Conclusions}

Alternative cosmologies with a quintessence component (dark
energy) may provide an explanation for the present accelerated
stage of the universe as suggested by the SNe Ia results. In this
work we have focused our attention on a possible dark energy
candidate, the so-called Chaplygin gas. The equation of state of
this dark energy component has been constrained by combining
Chandra observations of the X-ray luminosity of galaxy clusters
and independent measurements of the Hubble parameter and of the
baryonic matter density as well as from a
statistical combination between X-ray data and recent SNe Ia
observations. We
have shown that stringent constraints on the free parameters of
the model, namely $A_s$, $\alpha$ and $\Omega_m$, can be obtained
from this combination of observational data.

It is also interesting to compare the results derived here with
another independent analyses. For example, using only SNe Ia data,
Fabris et al. \cite{fabris} found $A_s = 0.93^{+0.07}_{-0.20}$ for
the original C-gas model ($\alpha = 1$) with the matter density
parameter constrained by the interval $0 \leq \Omega_{\rm{m}} \leq
0.35$. The same analysis for $\Omega_{\rm{m}} = \Omega_{\rm{b}} =
0.04$ (in which the C-gas plays the role of both dark matter and
dark energy) provides $A_s = 0.87^{+0.13}_{-0.18}$. These values
agree at some level with the ones obtained from statistics of
gravitational lensing (SGL), i.e., $A_s \geq 0.72$ \cite{dev} and
age estimates of high-$z$ galaxies (OHRG's), $A_s \geq 0.85$ -
$A_s \geq 0.99$ for the interval $\Omega_{\rm{m}} =0.2 - 0.4$ with
lower values of $A_s$ corresponding to lower $\Omega_{\rm{m}}$
\cite{jailson}. The original Chaplygin gas model, however, seems
to be incompatible with the localization of the acoustic peak of
CMB as given by WMAP \cite{benn} and BOOMERANG \cite{bern} data.
For the case of a generalized component, the same analysis shows
that for intermediary values of the spectral tilt $n_s$, the C-gas
model is favored by this data set if $\alpha \simeq 0.2$
\cite{bert1}. A similar analysis for BOOMERANG and Arqueops
\cite{benoi} data implies $0.57 \leq A_s \leq 0.91$ for $\alpha
\leq 1$ \cite{bento} whereas an investigation involving WMAP and
SNe Ia data sets restricts $\alpha$ to be $0 \leq \alpha \leq 0.2$
\cite{finelli}.

It should be stressed that our results are in line with the above
quoted independent studies. In particular, even considering that
the parameter $A_s$ is quite insensitive to the X-ray data alone,
the matter density parameter is very well constrained. This result
is also in agreement with the limits derived by Allen et
al. \cite{allen1} for $\Lambda$CDM models. In addition, as shown in
Figure 3b, by combining the X-ray and SNe Ia data sets, more
stringent constraints on the parameters $\Omega_{\rm{m}}$ and
$A_s$ are readily obtained. From the above analyses we also note that the $\alpha$
parameter is more strongly restricted if causality requirements
($v_s^{2} \leq 1)$ are imposed (see Figure 4). However, it seems that a even better
method to place limits on such a parameter is through the physics of the perturbations,
i.e., CMB and LSS data (see, e.g., \cite{stat1,fin}).

\begin{acknowledgments}
The authors are supported by the Conselho Nacional de
Desenvolvimento Cient\'{\i}fico e Tecnol\'{o}gico (CNPq - Brasil)
and CNPq (62.0053/01-1-PADCT III/Milenio).
\end{acknowledgments}


\end{document}